\documentclass[aps,pra, twocolumn,showpacs,superscriptaddress]{revtex4-1}

\usepackage{graphicx}
\usepackage{amsmath, calc}
\usepackage{amsfonts}
\usepackage{amssymb}
\usepackage{bm}
\usepackage{bbm}
\usepackage{color}
\usepackage{array}
\usepackage{times}
\usepackage{units}
\usepackage{comment}

\usepackage[pdftex, colorlinks=true, linkcolor=blue, citecolor=blue, urlcolor=blue]{hyperref}

\begin{document}

\title{Radiative frequency shifts in nanoplasmonic dimers}

\author{Charles A. Downing}
\affiliation{Universit\'{e} de Strasbourg, CNRS, Institut de Physique et Chimie des Mat\'{e}riaux de Strasbourg,  UMR 7504, F-67000 Strasbourg, France}

\author{Eros Mariani}
\affiliation{School of Physics and Astronomy, University of Exeter, Stocker Rd.\ EX4 4QL Exeter, UK}

\author{Guillaume Weick}
\email{guillaume.weick@ipcms.unistra.fr} 
\affiliation{Universit\'{e} de Strasbourg, CNRS, Institut de Physique et Chimie des Mat\'{e}riaux de Strasbourg,  UMR 7504, F-67000 Strasbourg, France}


\begin{abstract}
We study the effect of the electromagnetic environment on the resonance frequency of plasmonic excitations in dimers of interacting metallic nanoparticles. 
The coupling between plasmons and vacuum electromagnetic fluctuations induces a shift in the resonance frequencies---analogous to the Lamb shift in atomic physics---which is usually not measurable in an isolated nanoparticle. In contrast, we show that this shift leads to sizable corrections to the level splitting induced by dipolar interactions in nanoparticle dimers. For the system parameters which we consider in this work, the ratio between the level splitting for the longitudinal and transverse hybridized modes takes a universal form dependent only on the interparticle distance and thus is highly insensitive to the precise fabrication details of the two nanoparticles. We discuss the possibility to successfully perform the proposed measurement using state-of-the-art nanoplasmonic architectures.  
\end{abstract}

\maketitle

\section{Introduction}
The classical damped harmonic oscillator offers a paradigmatic example of the effect of 
the environment on the dynamics of a system. In the standard textbook case of a linear oscillator, 
the presence of damping forces proportional to its velocity yields a broadening of the resonance 
spectrum as well as a shift of its resonance frequency \cite{landau}. The quantum-mechanical analog of this 
phenomenon has been studied in the context of atomic physics by analyzing the effects of
the electromagnetic environment (i.e., photons) onto the atomic energy levels \cite{cohen}. 
The spontaneous emission of photons by electrons in an atom results in a finite broadening of their 
energy levels. In parallel, the quantum fluctuations of the electromagnetic vacuum lead to a 
renormalization of the atomic energy levels \cite{Milonni1994}. This Lamb shift \cite{Lamb1947, Bethe1947} 
is crucial for addressing 
the experimentally-observed atomic spectra, and represents a milestone result in early quantum
electrodynamics.

Several key concepts of atomic physics have been observed in ``artificial atoms'' consisting of metallic 
clusters  \cite{deHeer1993, Brack1993}.
The collective oscillation of valence electrons in a metallic nanoparticle---called a
localized surface plasmon (LSP)---has been successfully modeled as a dipolar oscillator whose harmonic response is
highly sensitive to the local environment \cite{Kreibig1995}. This sensitivity is exploited 
in a variety of devices, such as biological \cite{Anker2008} and chemical sensors \cite{Mayer2011}, as well as in plasmon-enhanced Raman scattering, where spatial resolution down to the single-molecule level has been recently achieved~\cite{Zhang2013}. 
LSP resonances are well known to exhibit a linewidth (which gives access to the lifetime of the collective modes)
that is crucially affected by different mechanisms involving the coupling between the electronic degrees of freedom and 
their environment \cite{bertsch}. Among these effects, radiation damping due to the emission of photons from individual nanoparticles leads 
to a line broadening proportional to their volume, which has been extensively studied in the literature~\cite{Kreibig1995}.
The corresponding small shift in the resonance frequency of individual LSPs, analogous to the Lamb shift in atomic physics, 
is however an elusive quantity to detect as it usually competes with other effects including, e.g., the spill-out of the electrons outside of the nanoparticle \cite{Brack1993} and retardation effects associated with multipolar excitations \cite{Kreibig1995}.

\begin{figure}[tb]
\includegraphics[width=\columnwidth]{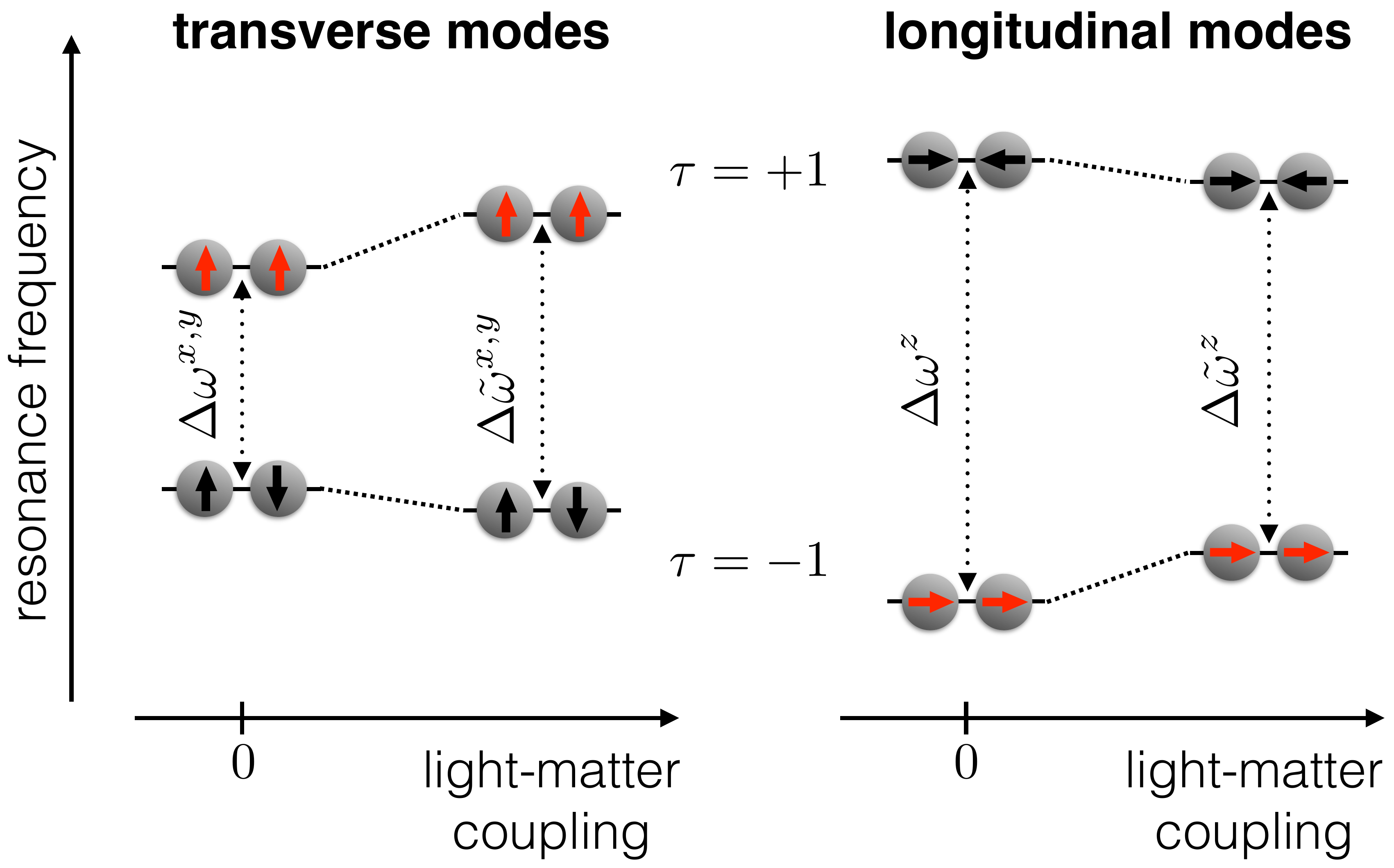}
\caption{Qualitative sketch of the effect of the photonic environment on the resonance frequencies of 
 hybridized bright (red arrows) and dark (black arrows) modes in metallic nanoparticle dimers.
 While the bright modes are blueshifted due to their interaction with the photonic environment, the dark modes, which are weakly coupled to light, are slightly redshifted. 
 This results in an increase (decrease) of the splitting between hybridized plasmonic modes for the transverse (longitudinal) polarization.} 
\label{fig:sketch} 
\end{figure}

In this work we show that the radiative frequency shift induced by the photonic environment can be unambiguously 
detected in a metallic nanoparticle dimer. Such a system hosts bright and dark hybridized collective modes (see Fig.\ \ref{fig:sketch}) resulting from the dipolar 
interaction between the LSPs in individual nanoparticles \cite{jain10_CPL}. 
These modes have been extensively studied both theoretically 
\cite{ruppi82_PRB, gerar83_PRB, nordl04_NL, dahme07_NL, bache08_PRL, zuloa09_NL, esteb12_NatureComm, zhang14_PRB, Brandstetter2015}
and experimentally~\cite{tamar02_APL, rechb03_OC, danck07_PRL, olk08_NL, chu09_NL, koh09_ACS, Barrow2014}. In particular, the dark plasmonic modes in a dimer have been detected by means of electron energy loss spectroscopy (EELS) \cite{chu09_NL, koh09_ACS, Barrow2014}.
Here we demonstrate that the energy splitting between the bright and dark hybridized modes is highly sensitive to the photonic environment, 
while being essentially unaffected by fluctuations in the individual LSP resonance frequency. 
More specifically, by analyzing the ratio between the splitting in the longitudinal and transverse modes, we unveil 
a deviation from the result obtained in the absence of the electromagnetic environment, which essentially only depends on 
the interparticle distance in the dimer. 
We propose an experimental methodology to detect this frequency shift and analyze its feasibility 
with current state-of-the-art techniques.

Our paper is organized as follows: in Sec.\ \ref{sec:1NP} we start by analyzing the radiative frequency shift in a single nanoparticle. 
In Sec.\ \ref{sec:dimer} we consider the case of a nanoplasmonic dimer, which constitutes the central focus of this work.
We discuss the experimental observability of the effect that we predict in Sec.\ \ref{sec:experiments} and conclude in Sec.\ \ref{sec:ccl}.
Details of our calculations are relegated to the Appendixes.

\section{Radiative frequency shift in a single plasmonic nanoparticle} 
\label{sec:1NP}
Before analyzing the nanoparticle dimer, we start by 
briefly discussing
the radiative frequency shift in a single plasmonic nanoparticle. Specifically, we consider 
a spherical metallic nanoparticle with radius $a$ containing $N_\mathrm{e}$ valence electrons. The nanoparticle supports three degenerate orthogonal LSPs with resonance frequency $\omega_0$ and polarizations $\sigma=x,y,z$. For alkaline nanoparticles, and ignoring the spill-out effect which is only relevant for very small nanoparticles (with $a\lesssim\unit[5]{nm}$) \cite{Kreibig1995}, $\omega_0$ is equivalent to the Mie frequency $(N_\mathrm{e} e^2/m_{\mathrm{e}} a^3)^{1/2}$, with $e$ and $m_\mathrm{e}$ denoting the electron charge and mass, respectively. 

The total Hamiltonian of the system is 
$H = H_{\mathrm{pl}} + H_\mathrm{ph} + H_{\mathrm{pl}\textrm{-}\mathrm{ph}}$.
Here, the Hamiltonian 
\begin{equation}
H_{\mathrm{pl}} = \hbar \omega_0 \sum_{\sigma=x,y,z}  {b^{\sigma}}^{\dagger} b^{\sigma}
\end{equation}
describes the three LSP modes,
where the bosonic operator $b^{\sigma}$ acts on an eigenstate $|n^\sigma\rangle$ by annihilating an LSP with polarization $\sigma$ 
as $b^\sigma|n^\sigma\rangle=\sqrt{n^\sigma}|n^\sigma-1\rangle$ (here, $n^\sigma$ is a non-negative integer). The term 
\begin{equation}
\label{eq:H_ph}
H_{\mathrm{ph}} = \sum_{\mathbf{k}, \hat{\lambda}_{\mathbf{k}}} \hbar \omega_{\mathbf{k}} {a_{\mathbf{k}}^{\hat{\lambda}_{\mathbf{k}}}}^{\dagger} a_{\mathbf{k}}^{\hat{\lambda}_{\mathbf{k}}}
\end{equation}
corresponds to vacuum photonic modes in a volume $\mathcal{V}$ with dispersion $\omega_{\mathbf{k}} = c |\mathbf{k}|$ ($c$ is the speed of light in vacuum),  
 where $a_{\mathbf{k}}^{\hat{\lambda}_{\mathbf{k}}}$ (${a_{\mathbf{k}}^{\hat{\lambda}_{\mathbf{k}}}}^{\dagger}$) annihilates (creates) a photon with 
 wavevector $\mathbf{k}$ and transverse polarization 
 $\hat{\lambda}_{\mathbf{k}}$ (with $\mathbf{k}\cdot\hat{\lambda}_{\mathbf{k}}=0$).
 Here and in what follows, hats designate unit vectors. 
In the long-wavelength limit ($k_0a\ll1$, with $k_0=\omega_0/c$), the plasmon-photon minimal coupling Hamiltonian in the Coulomb gauge reads \cite{craig, Milonni1994}
\begin{equation}
\label{eq:plphsingle}
 H_{\mathrm{pl}\textrm{-}\mathrm{ph}} = \frac{e}{m_{\mathrm{e}}} \mathbf{\Pi} \cdot \mathbf{A} (\mathbf{R}) + \frac{N_{\mathrm{e}} e^2}{2 m_{\mathrm{e}}} \mathbf{A}^2 (\mathbf{R}),
\end{equation}
where 
\begin{equation}
\mathbf{\Pi}=\mathrm{i}\sqrt{\frac{N_\mathrm{e}m_\mathrm{e}\hbar\omega_0}{2}}\sum_{\sigma=x,y,z}
\left({b^\sigma}^\dagger-b^\sigma\right)\hat\sigma
\end{equation}
 is the LSP momentum, 
\begin{equation}
\label{eq:A}
\mathbf{A}(\mathbf{r})=\sum_{\mathbf{k}, \hat\lambda_{\mathbf{k}}}
\hat\lambda_{\mathbf{k}}
\sqrt{\frac{2\pi\hbar}{\mathcal{V}\omega_\mathbf{k}}}
\left(
a_\mathbf{k}^{\hat\lambda_{\mathbf{k}}}\mathrm{e}^{\mathrm{i}\mathbf{k}\cdot\mathbf{r}}
+{a_\mathbf{k}^{\hat\lambda_{\mathbf{k}}}}^\dagger\mathrm{e}^{-\mathrm{i}\mathbf{k}\cdot\mathbf{r}}
\right)
\end{equation}
is the vector potential, and $\mathbf{R}$ is the location of the center of the nanoparticle.
In what follows, we disregard the coupling of the LSPs to electron-hole pairs (which leads to Landau damping \cite{Kawabata1966, bertsch, Kreibig1995, Weick2005} as well as to a small renormalization of the resonance frequency \cite{Weick2006}) which is only relevant for tiny nanoparticles ($a\lesssim\unit[5]{nm}$).

Treating the coupling Hamiltonian \eqref{eq:plphsingle} up to second order in perturbation theory,
we calculate the corrections to 
the plasmonic energy levels $E_{n^\sigma} = E_{n^\sigma}^{(0)} + E_{n^\sigma}^{(1)} + E_{n^\sigma}^{(2)}$, where the unperturbed contribution is $E_{n^\sigma}^{(0)} = n^{\sigma}  \hbar\omega_0$. 
The first-order correction 
\begin{equation}
\label{eq:E1}
E_{n^\sigma}^{(1)} =2\pi\hbar\omega_0^2\frac{a^3}{\mathcal{V}}\sum_\mathbf{k}\frac{1}{\omega_\mathbf{k}}
\end{equation}
stems from the second term in the right-hand side of Eq.\ \eqref{eq:plphsingle} and 
corresponds to 
a global energy shift that does not depend on the quantum number $n^\sigma$ and therefore  does not contribute to a modification of the LSP resonance frequency. 
The second-order correction  
\begin{equation}
\label{eq:secondorder}
 E_{n^\sigma}^{(2)} = \pi \hbar\omega_0^3 \frac{a^3}{\mathcal{V}} \sum_{\mathbf{k}, \hat{\lambda}_{\mathbf{k}}} \frac{|\hat{\sigma} \cdot \hat{\lambda}_{\mathbf{k}} |^2}{\omega_{\mathbf{k}}} \frac{ (2 n^{\sigma} + 1) \omega_{\mathbf{k}}-\omega_0}{\omega_0^2 - \omega_{\mathbf{k}}^2}
\end{equation}
is associated with the first term in the right-hand side of Eq.~\eqref{eq:plphsingle} and 
arises due to the emission and reabsorption of virtual photons by the plasmonic state $|n^\sigma\rangle$. 
In the expression above, the summation excludes the term for which $\omega_{\mathbf{k}}=\omega_0$. 

\begin{figure}[tb]
\includegraphics[width=\columnwidth]{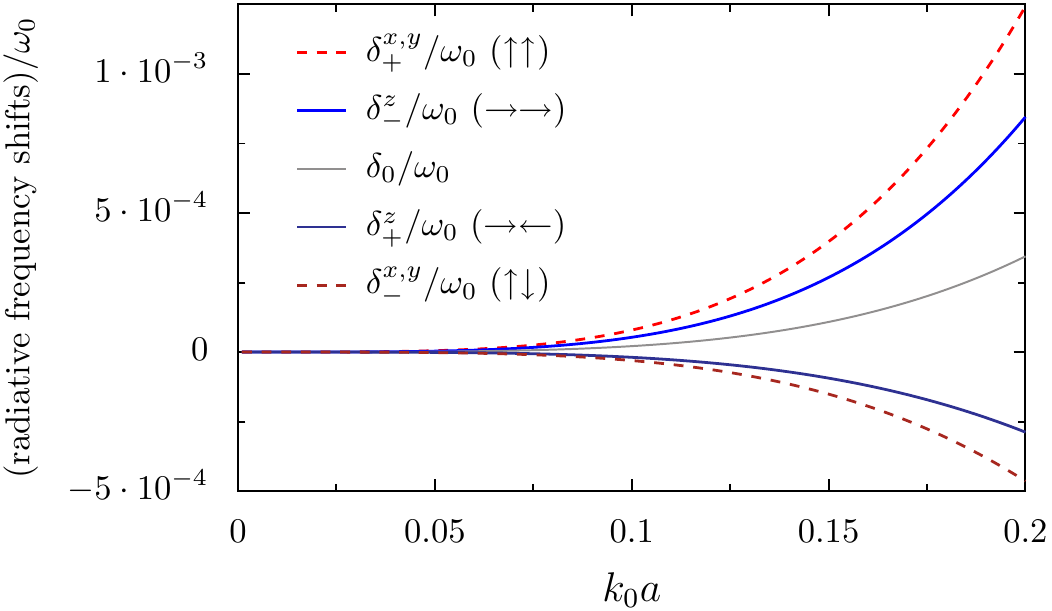}
\caption{
Radiative frequency shifts (in units of the bare LSP resonance frequency $\omega_0$) as a function of the (reduced) nanoparticle size $k_0a$.
Thin gray line: $\delta_0$ for an isolated nanoparticle [cf.\ Eq.\ \eqref{eq:shiftsingle}]. 
Colored lines: $\delta_\tau^\sigma$ for a dimer with interparticle distance $d=3a$ [cf.\ Eq.\ \eqref{eq:shiftdimer}]. 
In the figure, $\omega_\mathrm{c}=c/a$.} 
\label{fig:delta} 
\end{figure}

The second-order correction \eqref{eq:secondorder} appears to be linearly divergent. Such a divergence can be regularized by following a renormalization procedure analogous to that originally used by Bethe in his analysis of 
the Lamb shift in atomic physics~\cite{Bethe1947, Milonni1994}. 
To second order in perturbation theory, the renormalized frequency difference between successive plasmonic energy levels $\tilde\omega_0 = (E_{n^{\sigma}+1} -E_{n^{\sigma}})/\hbar$ is then independent of the quantum number $n^{\sigma}$ and reads $\tilde\omega_0 = \omega_0 + \delta_0 $, where the radiative frequency shift is given by (see Appendix~\ref{app:1NP} for details)
\begin{equation}
\label{eq:shiftsingle}
\delta_0=\frac{\omega_0}{3 \pi}\left( k_0 a\right)^3 \ln{ \left( \frac{ \omega_\mathrm{c} / \omega_0 + 1 }{\omega_\mathrm{c} / \omega_0 - 1} \right)}.
\end{equation}
Here, $\omega_\mathrm{c}$ is an ultraviolet cutoff of the order of $c/a$, which corresponds to the wavelength below which the dipolar approximation used in Eq.\ \eqref{eq:plphsingle} breaks down.
With this choice of cutoff we have $\omega_\mathrm{c} / \omega_0 = 1/k_0 a>1$, such that Eq.\ \eqref{eq:shiftsingle} corresponds to a blueshift ($\delta_0 > 0$) of the LSP resonance frequency, in qualitative agreement with the result for the Lamb shift in atomic physics \cite{Bethe1947, Milonni1994, welto48_PR}. 
Furthermore, this radiative frequency shift increases for increasing nanoparticle size, as shown by the thin gray line in Fig.\ \ref{fig:delta}. In the limit $k_0 a \ll 1$, the single-particle radiative shift is approximately given by 
$\delta_0 \simeq 2\omega_0 (k_0 a)^4/3\pi$.
 We point out that the shift $\delta_0$ is challenging to measure experimentally as it may be masked by other mechanisms leading to a renormalization of the Mie frequency $\omega_0$, such as retardation effects which become prominent for larger nanoparticles \cite{Kreibig1995}. 
In stark contrast, the radiative frequency shift can be detected by analyzing the spectrum of a nanoparticle dimer, as we show below.

\section{Radiative frequency shifts in nanoparticle dimers} 
\label{sec:dimer}
\begin{figure}[tb]
 \includegraphics[width=.85\columnwidth]{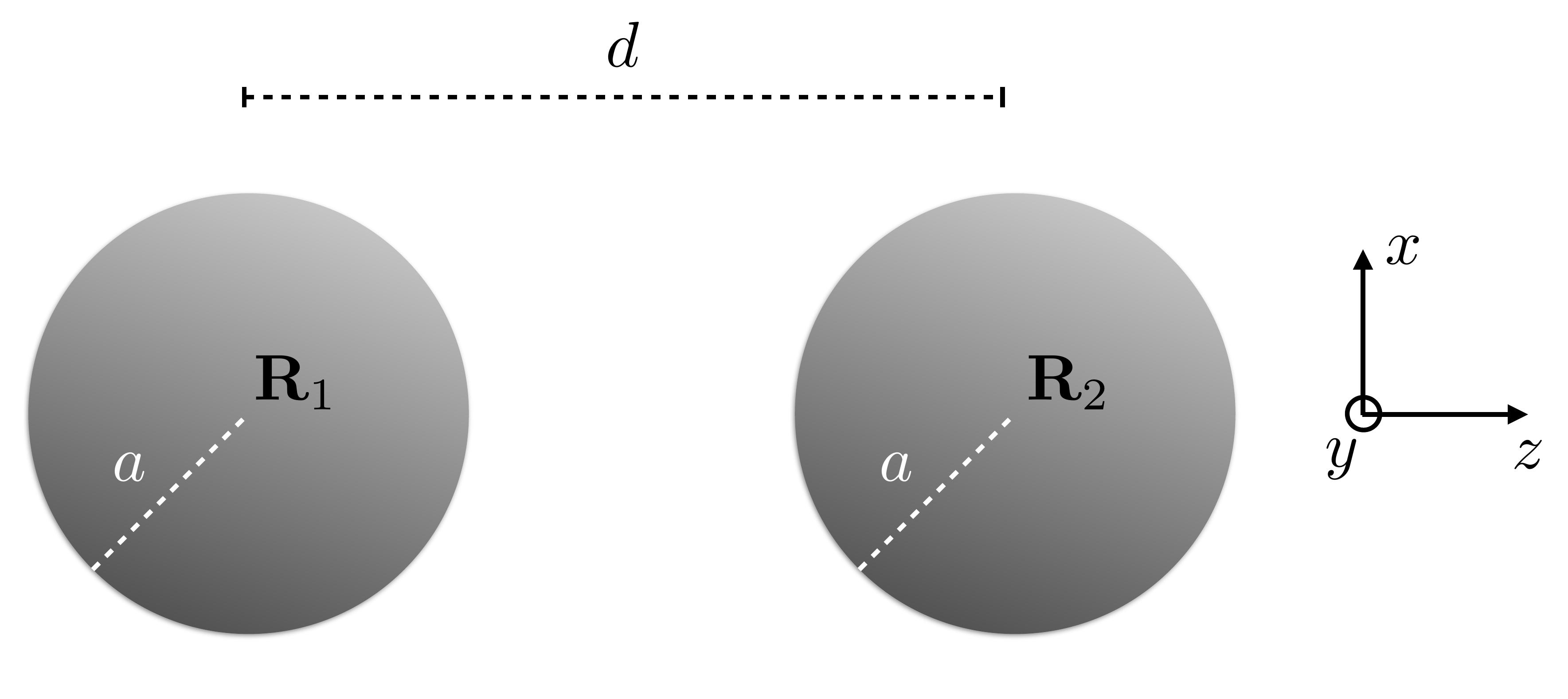}
 \caption{Sketch of a plasmonic dimer composed of two identical spherical metallic nanoparticles of radius $a$, separated by a center-to-center distance $d$. The location of the nanoparticle centers are denoted by $\mathbf{R}_1$ and $\mathbf{R}_2$, respectively.}
 \label{fig:sketch_dimer}
\end{figure}

We now consider a dimer formed by two identical spherical metallic nanoparticles of radius $a$
with a center-to-center distance $d$ along the $z$ direction (see Fig.\ \ref{fig:sketch_dimer}). In the regime $3a \lesssim d \ll k_0^{-1}$ \cite{Park2004}, the near-field quasistatic dipole-dipole interaction between the LSPs in each nanoparticle (with resonance frequency $\omega_0$) results in hybridized plasmonic modes governed by the Hamiltonian 
\begin{equation}
\label{eq:pldimer}
H_{\mathrm{pl}} = \sum_{\sigma=x,y,z}\sum_{\tau=\pm} \hbar \omega_{\tau}^{\sigma} {B_{\tau}^{\sigma}}^{\dagger} B_{\tau}^{\sigma}.
\end{equation}
Here the bosonic Bogoliubov operator $B_{\tau}^{\sigma}$ (${B_{\tau}^{\sigma}}^{\dagger}$) 
annihilates (creates) a coupled plasmonic mode with polarization $\sigma$ 
(for the transverse modes $\sigma=x,y$; for the longitudinal ones $\sigma=z$). The 
eigenfrequencies read (see Refs.\ \cite{Brandstetter2015, Brandstetter2016} and Appendix \ref{app:dimer} for details)
\begin{equation}
\label{eq:omega_bare}
\omega_{\tau}^{\sigma} = \omega_0 \sqrt{1 + 2 \tau |\eta_{\sigma}| \frac{\Omega}{\omega_0}}.
\end{equation}
In the expression above, $\Omega = \omega_0 (a/d)^3 /2 \ll \omega_0$, $\eta_{x,y} = 1$ for the transverse modes, and  $\eta_{z} = -2$ for the longitudinal ones. 
The label $\tau$ distinguishes the high- ($\tau=+$) and low-energy ($\tau=-$) coupled plasmonic modes (see Fig.\ \ref{fig:sketch}). 
Importantly, the high-energy transverse and low-energy longitudinal excitations correspond to symmetric bright modes coupled to the photonic environment. Conversely, the low-energy transverse and high-energy longitudinal modes are dark antisymmetric modes weakly coupled to light. 
As we will demonstrate in what follows, the bright modes experience a blueshift due to the interaction with the photonic environment, while the dark modes are slightly redshifted, as sketched in Fig.\ \ref{fig:sketch}. 

The total Hamiltonian $H=H_{\mathrm{pl}} + H_\mathrm{ph} + H_{\mathrm{pl}\textrm{-}\mathrm{ph}}$ of the dimer now includes the plasmonic term \eqref{eq:pldimer}, the photonic one~\eqref{eq:H_ph}, and the coupling Hamiltonian $H_{\mathrm{pl}\textrm{-}\mathrm{ph}}$, which is easily generalized from Eq.\ \eqref{eq:plphsingle} to two nanoparticles located at $\mathbf{R}_1$ and $\mathbf{R}_2 = \mathbf{R}_1+ d\, \hat{z}$ [see Eq.\ \eqref{eqS:HamCoupling}]. 

The perturbative calculation of the radiative shifts proceeds along the line as that of a single nanoparticle (cf.\ Sec.\ \ref{sec:1NP}). 
As a result the renormalized frequency spectrum of the dimer reads 
$\tilde\omega_{\tau}^{\sigma} =\omega_{\tau}^{\sigma} + \delta_{\tau}^{\sigma} $, where the radiative frequency shift is given by
(see Appendix \ref{app:dimer} for details)
\begin{equation}
\label{eq:shiftdimer}
\delta_{\tau}^{\sigma} = \frac{{\omega_{\tau}^{\sigma}}^2}{3 \pi\omega_0}\left( k_0 a \right)^3 
\ln{ \left( \frac{ \omega_\mathrm{c} / \omega_{\tau}^{\sigma} + 1 }{\omega_\mathrm{c} / \omega_{\tau}^{\sigma} - 1} \right)} + \tau
\frac{ |\eta_{\sigma}|\omega_0\Omega}{\pi\omega_{\tau}^{\sigma}}   g_{\tau}^{\sigma},
\end{equation}
where 
\begin{align}
\label{eq:superlong}
 g_{\tau}^{\sigma} =&\; 
 \left( \frac{k_{\tau}^{\sigma}}{k_\mathrm{c}} \right)^2 
 \left[ \sin \left(k_\mathrm{c} d\right) - k_\mathrm{c} d\cos \left(k_\mathrm{c} d \right) \right] \nonumber\\
 &- \left[ 2 f_{\tau}^{\sigma} + \left(k_\tau^\sigma d\right)^2\right] \mathrm{Si} \left(k_\mathrm{c} d\right) \nonumber\\
  &+ \sum_{\zeta=\pm}\left\{
  r_{\tau}^{\sigma}\mathrm{Si} \left( [ k_\mathrm{c} + \zeta k_{\tau}^{\sigma} ] d\right) 
  - \zeta s_{\tau}^{\sigma} \mathrm{Ci} \left( [ k_\mathrm{c} + \zeta k_{\tau}^{\sigma} ] d \right) \right\}.
\end{align}
In Eq.\ \eqref{eq:superlong}, 
$\mathrm{Si}(z)$ and $\mathrm{Ci}(z)$ denote the sine and cosine integrals, while
\begin{equation}
f_{\tau}^{\sigma}= 1 -  \frac 12\left( 1 + \mathrm{sgn} \{ \eta_{\sigma} \} \right) ( k_{\tau}^{\sigma} d )^2,
\end{equation}
\begin{equation}
r_{\tau}^{\sigma} =  f_{\tau}^{\sigma} \cos \left(k_{\tau}^{\sigma} d \right) + k_{\tau}^{\sigma} d \sin \left( k_{\tau}^{\sigma} d \right),
\end{equation}
and
\begin{equation}
s_{\tau}^{\sigma} =  f_{\tau}^{\sigma} \sin \left( k_{\tau}^{\sigma} d\right) - k_{\tau}^{\sigma} d\cos \left( k_{\tau}^{\sigma} d\right), 
\end{equation}
with $k_{\tau}^{\sigma}=\omega_{\tau}^{\sigma}/c$ and $k_\mathrm{c}=\omega_\mathrm{c}/c$.

The radiative shifts $\delta_\tau^\sigma$ from Eq.\ \eqref{eq:shiftdimer} are plotted in Fig.\ \ref{fig:delta} as a function of the nanoparticle size for an interparticle distance $d=3a$. As can be seen from the figure, both the bright transverse (red dashed line) and longitudinal (solid light blue line) hybridized modes experience a blueshift due to the coupling to the photonic environment, whereas the dark modes are slightly redshifted (solid dark blue and dashed brown lines in Fig.\ \ref{fig:delta}). 
The bright modes experience an enhanced frequency shift compared to the single nanoparticle result (solid gray line), similar to the cooperative Lamb shift in many-atom systems \cite{scull09_PRL, rohls10_Science}.
We have verified that in the limit of large interparticle distance ($d\gg a$), all radiative shifts \eqref{eq:shiftdimer} asymptotically tend to the single-particle result \eqref{eq:shiftsingle}.

\begin{figure}[tb]
\includegraphics[width=\columnwidth]{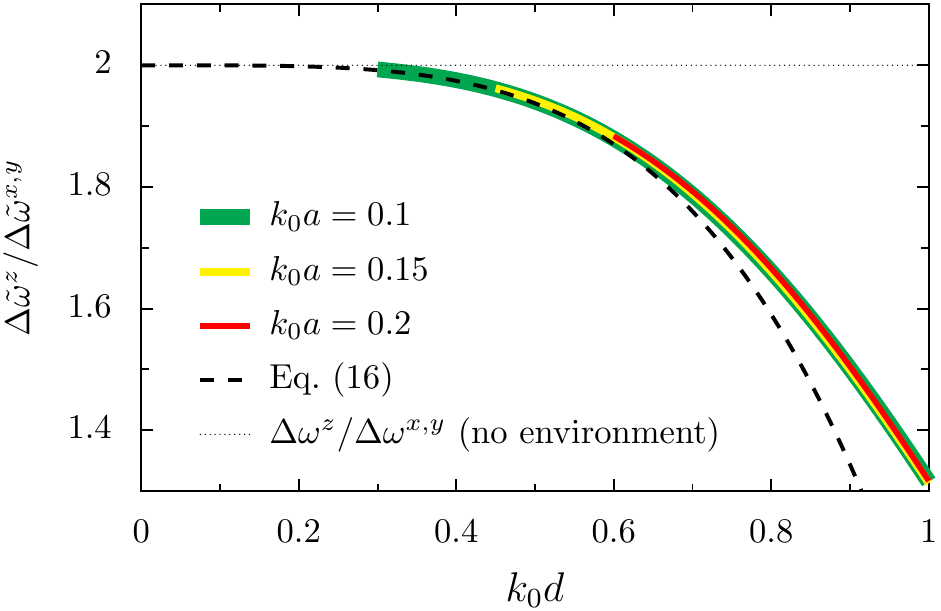}
\caption{
Colored solid lines: ratio $\Delta\tilde\omega^z/\Delta\tilde\omega^{x,y}$ in a nanoparticle dimer as a function of the interparticle distance $d$ for 
 increasing nanoparticle radii $a$ as obtained from Eq.\ \eqref{eq:shiftdimer}. 
 Dashed line: approximate result from Eq.\ \eqref{eq:beauty}. Dotted line: ratio $\Delta\omega^z/\Delta\omega^{x,y}$ in the absence of the coupling with the photonic environment.
In the figure, $\omega_\mathrm{c}=c/a$.} 
\label{fig:universality} 
\end{figure}

\begin{figure*}[tb]
\includegraphics[width=\linewidth]{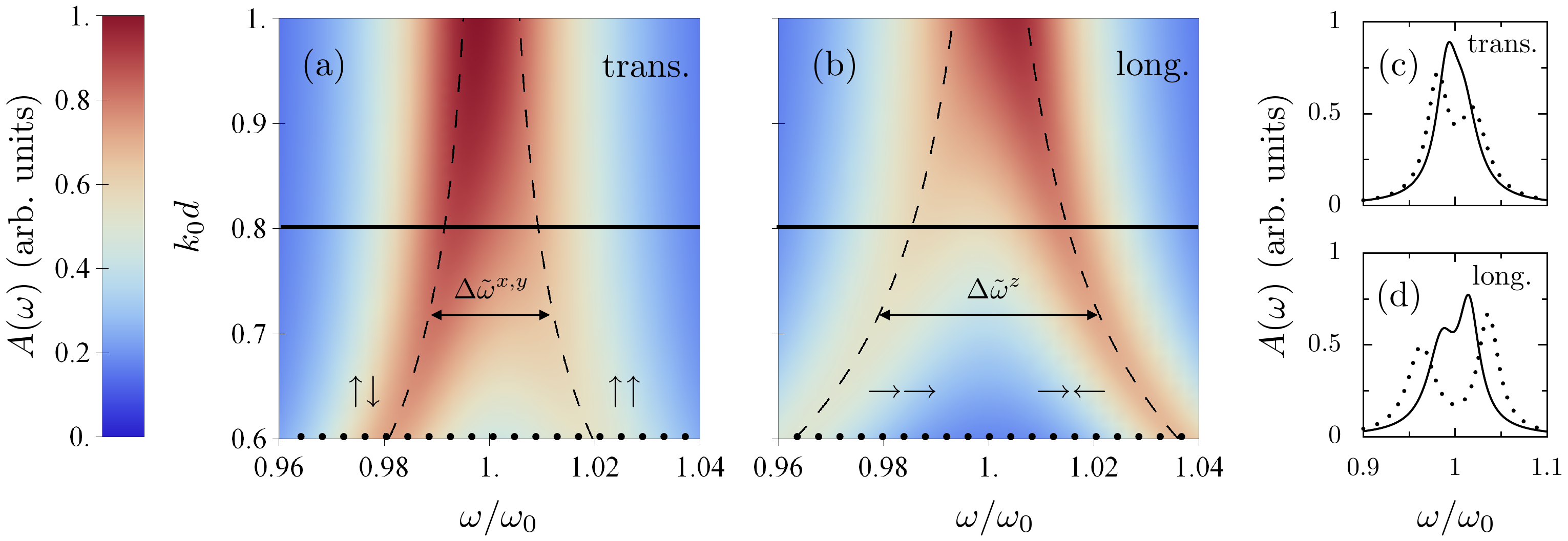}
\caption{
(a),(b) Absorption spectrum of the dimer $A(\omega)$ as a function of frequency $\omega$ and interparticle separation $d$ for the (a) transverse and (b) longitudinal polarizations.
The dashed lines indicate the location of the resonance frequencies $\tilde\omega_\tau^\sigma$ of the hybridized plasmonic modes.
(c),(d) Cuts of the absorption spectrum at $k_0d=0.6$ (dotted lines) and $k_0d=0.8$ (solid lines) for the (c) transverse and (d) longitudinal polarizations, 
corresponding to the dotted and solid lines in panels (a) and (b), respectively.
In the figure, $k_0a=0.2$ and $\gamma^{\text{O}} = 0.027 \omega_0$, corresponding to a dimer of Ag nanoparticles with $a=\unit[15]{nm}$ and $\omega_0=\unit[2.6]{eV/\hbar}$.} 
\label{fig:observability} 
\end{figure*}

The radiative shifts \eqref{eq:shiftdimer}  produce a renormalization of the plasmonic resonances in analogy with the single-particle case. 
These shifts are however difficult to detect in experiments as they are masked by additional size-dependent fluctuations in the 
LSP resonance frequencies. The presence of hybridized states in a dimer offers a way out of this problem by looking at the 
effect of the environment on the frequency splitting $\Delta\tilde\omega^\sigma=\tilde \omega_+^\sigma-\tilde \omega_-^\sigma$ 
between bright and dark modes. The most informative quantity revealing this effect is the dimensionless ratio 
$\Delta\tilde\omega^z/\Delta\tilde\omega^{x,y}$. 
In the absence of the electromagnetic environment, $\Delta\omega^z/\Delta\omega^{x,y}=2$ 
(with $\Delta\omega^\sigma= \omega_+^\sigma- \omega_-^\sigma$) is independent of the interparticle separation, up to quadratic corrections in $\Omega/\omega_0\ll1$. 
As sketched in Fig.\ \ref{fig:sketch}, the electromagnetic environment has opposite effects on the transverse and longitudinal modes, leading to an 
increase/decrease of the frequency splitting, respectively. 
This induces a pronounced deviation of $\Delta\tilde\omega^z/\Delta\tilde\omega^{x,y}$ with respect to $\Delta\omega^z/\Delta\omega^{x,y}$, as shown in
Fig.\ \ref{fig:universality} where such a ratio is plotted as a function of the interparticle distance $d$ for different 
nanoparticle radii $a$ (solid colored lines in the figure). 
To be consistent with our dipolar approximation \cite{Park2004}, each colored line starts at $d=3a$.

It is apparent from Fig.\ \ref{fig:universality} that our results for $\Delta\tilde\omega^z/\Delta\tilde\omega^{x,y}$ are essentially independent of the nanoparticle radius $a$, despite the clear size dependence of the radiative frequency shifts for individual dimer levels [cf.\ Eq.\ \eqref{eq:shiftdimer} and Fig.\ \ref{fig:delta}], and 
sensitive to the interparticle separation $d$ only.
This unexpected result can be understood by performing a systematic expansion of 
Eq.\ \eqref{eq:shiftdimer} 
in $\Omega/\omega_0\ll1$ with $k_0a\ll1$ and $k_0d\ll1$, leading to the remarkably simple approximate expression
(see Appendix \ref{app:beauty} for details)
\begin{equation}
\label{eq:beauty}
\frac{\Delta\tilde\omega^z}{\Delta\tilde\omega^{x,y}}\simeq2-\left(k_0d\right)^4. 
\end{equation}
As can be seen from the dashed line in Fig.\ \ref{fig:universality}, such a result is in excellent agreement with the exact expression, even when 
$k_0d$ is a significant fraction of unity. 

A legitimate question to address here is the robustness of the universal scaling found in the expression above against multipolar interactions of the LSPs with 
vacuum electromagnetic modes beyond the dipolar approximation in Eq.\ \eqref{eq:plphsingle}. Since, to leading order in $k_0a\ll1$, the radiative shifts in Eqs.\ 
\eqref{eq:shiftsingle} and \eqref{eq:shiftdimer} scale as $(k_0a)^4$, it is to be expected that the next leading-order correction to these results goes, at least, as $(k_0a)^5$. 
This would yield an additional nonuniversal, $a$-dependent contribution of the order of $(k_0a)^2 (k_0d)^3$ to the ratio $\Delta\tilde\omega^z/\Delta\tilde\omega^{x,y}$ given in Eq.\ \eqref{eq:beauty} (see Appendix \ref{app:beauty}). Thus such nonuniversal corrections can be safely neglected as long as $(k_0a)^2\ll k_0d$, which is the case for the parameters used in Fig.\ \ref{fig:universality}.
We have further checked that Eq.~\eqref{eq:beauty} remains valid for a heterogeneous dimer where the two LSP resonance frequencies differ by $\delta\omega$, 
as long as $\delta\omega\lesssim\Omega$ (see Appendix \ref{app:hetero}). 
This may be key for the experimental detection of our predicted effect, as it 
is essentially insensitive to the precise size of the nanoparticles in the dimer and their respective resonance frequencies, and hence to the challenges involved in the nanofabrication of the sample.

\section{Experimental detection} 
\label{sec:experiments}
Our proposal first requires both the excitation and detection of the dark and bright plasmonic modes. 
While the latter can be readily accessed optically, the former requires the use of the EELS technique, 
which has recently achieved a remarkable resolution of the order of $\unit[10]{meV}$ \cite{kriva14_Nature}.
Dark plasmonic modes in nanoparticle dimers \cite{chu09_NL, koh09_ACS, Barrow2014}
have already been successfully resolved using EELS. 
Alternatively, one may excite both bright and dark modes by optically driving one of the nanoparticles of the dimer only, 
or by using twisted light \cite{kerbe17_ACSPhoton}.

The frequency splitting between bright and dark modes $\Delta\tilde\omega^\sigma$ can then be resolved as long as it is not 
significantly smaller than the linewidth $\gamma^\sigma_\tau=\gamma^\mathrm{O}+\gamma^{\sigma, \mathrm{r}}_\tau$ of the respective resonances. 
The latter quantity involves both the nonradiative Ohmic damping characterized by the decay rate $\gamma^\mathrm{O}$, and 
the size-dependent radiative damping (which results from the decay of a collective plasmon into photons) with rate $\gamma^{\sigma, \mathrm{r}}_\tau$. An explicit expression for $\gamma^{\sigma, \mathrm{r}}_\tau$ can be found in Eq.\ (B12) of Ref.~\cite{Brandstetter2016} 
and presents an $a^3$ dependence. 
The Landau damping (i.e., the decay of the collective modes into electron-hole pairs \cite{Kreibig1995, bertsch, Brack1993})
has been evaluated for the case of a nanoparticle dimer in Refs.\ \cite{Brandstetter2015, Brandstetter2016} 
and scales as $1/a$. 
This contribution can be neglected for the nanoparticle sizes we consider in this work.

In Fig.\ \ref{fig:observability} we show the absorption spectrum $A(\omega)$ of the dimer system as a function of frequency $\omega$ and for increasing nanoparticle separation $d$ for the transverse [panel (a)] and longitudinal modes [panel (b)]. 
Panels (c) and (d) respectively correspond to cuts of Figs.\ \ref{fig:observability}(a) and \ref{fig:observability}(b), at $k_0d=0.6$ (dotted lines) and $0.8$ (solid lines).
The absorption spectrum $A(\omega)=\sum_{\tau=\pm}(\gamma_\tau^\sigma/2)/[(\omega-\tilde\omega_\tau^\sigma)^2+(\gamma_\tau^\sigma/2)^2]$ is assumed to be proportional, for a given polarization $\sigma$, to the sum of two Lorentzians 
centered at $\tilde\omega_\pm^\sigma$ and with full width at half maximum $\gamma_\pm^\sigma$.
In the figure, $k_0a=0.2$, which corresponds to Ag nanoparticles with Mie frequency $\omega_0=\unit[2.6]{eV/\hbar}$, radius $a=\unit[15]{nm}$, and interparticle separation ranging from 
$d=\unit[45]{nm}$ to $\unit[75]{nm}$. The value of the Ohmic decay rate is taken from experiments \cite{charl89_ZPD}.
As can be seen from Fig.~\ref{fig:observability}, the bright and dark modes can be well distinguished up to separations of the order of $k_0d\simeq0.8$ ($0.95$) for the transverse (longitudinal) polarization. 
Therefore, our proposal of detecting radiative shifts in a nanoparticle dimer, which relies on the ratio 
$\Delta\tilde\omega^z/\Delta\tilde\omega^{x,y}$ of the frequency splittings between bright and dark modes (see Fig.\ \ref{fig:universality}), is within experimental accessibility. Moreover, there is scope for observing the approximate quartic 
$d$-dependence of $\Delta\tilde\omega^z/\Delta\tilde\omega^{x,y}$ predicted by Eq.\ \eqref{eq:beauty}.
Control over the interparticle separation $d$ can be achieved with a variety of experimental setups, including 
the nanofabrication of different samples \cite{jain10_CPL}, the deposition of dimers on stretchable substrates \cite{Huang2010, Aksu2011, Cui2012}, or the use of a movable tip carrying one of the nanoparticles of the dimer
\cite{danck07_PRL, olk08_NL}.

\section{Conclusion} 
\label{sec:ccl}
We have proposed a dimer of metallic nanoparticles as an ideal testbed to observe the nanoplasmonic analog of the well-known Lamb shift in atomic physics. While a single nanoparticle has a radiative frequency shift that may be too difficult to measure experimentally at present, the same effect in a dimer, which is characterized by having both bright and dark modes, is within current experimental reach. 
For a dimer, we have shown that the resonance frequencies of the bright modes 
experience a blueshift due to the coupling to the photonic environment, while the dark modes are redshifted. 
As a result, the frequency splitting between bright and dark modes is larger (smaller) for the transverse (longitudinal) modes as
compared to those without coupling to the photonic environment. 
Remarkably, the ratio of splittings between the longitudinal and transverse polarizations has a 
universal
quartic dependence 
on the interparticle separation alone that would otherwise be absent if the plasmon-photon coupling was ignored. 
This prediction offers the tantalizing prospect to be measured in cutting-edge nanoplasmonic experiments.

\begin{acknowledgments}
We thank St\'ephane Berciaud, Fran\c{c}ois Gautier, Simon Horsley, and Guillaume Schull for stimulating discussions. 
This work was partially funded by the Agence Nationale de la Recherche (Project No.\ ANR-14-CE26-0005 Q-MetaMat), the Centre National de la Recherche Scientifique through the Projet International de Coop\'eration Scientifique program (Contract No.\ 6384 APAG), the Leverhulme Trust (Research Project Grant No.\ RPG-2015-101), and the Royal Society (International Exchange Grant No.\ IE140367, Newton Mobility Grants No.\  2016/R1 UK-Brazil, and Theo Murphy Award No.\ TM160190).
\end{acknowledgments}

\appendix
\section{Details of the calculation of the radiative shift in a single nanoparticle}
\label{app:1NP}
In this Appendix, we comment on the renormalization procedure for the second-order correction \eqref{eq:secondorder} to the plasmon 
energy levels in a single plasmonic nanoparticle, and provide details of the calculation of the resulting radiative frequency shift.

In the continuum limit where $\sum_{\mathbf{k}} \to {\mathcal{V}} \mathcal{P}\int \mathrm{d}^3 \mathbf{k}/{(2 \pi)^3}$ (here, $\mathcal{P}$ denotes the Cauchy principal value), the second-order correction \eqref{eq:secondorder} is linearly divergent. 
In order to regularize such a divergency, we follow Bethe's renormalization procedure in his analysis of the Lamb shift in atomic physics~\cite{Bethe1947, Milonni1994}.
First, we introduce an ultraviolet cutoff $k_\mathrm{c}$ of the order of $1/a$, which corresponds to the wavelength below which the dipolar approximation used in Eq.\ \eqref{eq:plphsingle} breaks down. 
Secondly, one subtracts from Eq.\ \eqref{eq:secondorder} the energy shift corresponding to free electrons with the same energy. This quantity is found by taking the limit of vanishing transition frequency in Eq.\ \eqref{eq:secondorder}, or equivalently  $\omega_0 \to 0$ inside the summation only \cite{Milonni1994}. Thus we arrive at the renormalized energy shift 
\begin{equation}
\label{renorm}
 \bar{E}_{n^\sigma}^{(2)} = \pi \hbar\omega_0^4 \frac{a^3}{\mathcal{V}}\sum_{\mathbf{k}, \hat{\lambda}_{\mathbf{k}}} \frac{|\hat{\sigma} \cdot \hat{\lambda}_{\mathbf{k}} |^2}{\omega_{\mathbf{k}}^2} \frac{ (2 n^{\sigma} + 1) \omega_0 - \omega_{\mathbf{k}}}{\omega_0^2 - \omega_{\mathbf{k}}^2},
\end{equation}
which is only logarithmically divergent with the cutoff $k_\mathrm{c}$, in analogy with the expression for the Lamb shift in atomic physics \cite{Bethe1947, Milonni1994}.
To second order in perturbation theory, the renormalized frequency difference between successive plasmonic energy levels then reads $\tilde\omega_0 = \omega_0 + \delta_0 $, where the radiative frequency shift is given by 
\begin{equation}
\label{eqS:lambsingle}
 \delta_0 = 2 \pi \omega_0^5 \frac{a^3}{\mathcal{V}} \sum_{\mathbf{k}, \hat{\lambda}_{\mathbf{k}}} \frac{|\hat{\sigma} \cdot \hat{\lambda}_{\mathbf{k}} |^2}{\omega_{\mathbf{k}}^2} \frac{1}{\omega_0^2 -\omega_{\mathbf{k}}^2 }.
\end{equation}

Carrying out the summation over photon polarization in Eq.~\eqref{eqS:lambsingle} via the relation 
\begin{equation}
\label{eqS:sum_pol}
\sum_{\hat{\lambda}_{\mathbf{k}}}{|\hat{\sigma} \cdot \hat{\lambda}_{\mathbf{k}} |}^2 = 1 - {( \hat{\sigma}\cdot {\hat{k}} )}^2
\end{equation}
and transforming the wavevector summation into a principal-value integral yields 
\begin{align}
\label{eqS:delta_0_inter}
\delta_0=&\;\frac{\omega_0^5a^3}{(2\pi c)^2}
\mathcal{P}\int_0^{k_\mathrm{c}}\frac{\mathrm{d}k}{\omega_0^2-(ck)^2}
\int_0^\pi\mathrm{d}\theta\sin{\theta}\nonumber\\
&\times
\int_0^{2\pi}\mathrm{d}\varphi[1-(\hat k\cdot\hat\sigma)^2].
\end{align}
The integral over $\varphi$ is easily evaluated,
\begin{equation}
\label{eqS:int_phi}
\int_0^{2\pi}\mathrm{d}\varphi[1-(\hat k\cdot\hat\sigma)^2]=
\pi|\eta_\sigma|\left(1+\mathrm{sgn}\{\eta_\sigma\}\cos^2{\theta}\right), 
\end{equation}
yielding for Eq.\ \eqref{eqS:delta_0_inter}
\begin{equation}
\delta_0=\frac{2\omega_0^5a^3}{3\pi c^3}\mathcal{P}\int_0^{\omega_\mathrm{c}}\frac{\mathrm{d}\omega}{\omega_0^2-\omega^2}, 
\end{equation}
with $\omega_\mathrm{c}=ck_\mathrm{c}$, where $\omega_\mathrm{c}>\omega_0$. The remaining integral over $\omega$ then gives the result 
of Eq.\ \eqref{eq:shiftsingle}.

\section{Radiative frequency shifts in homogeneous nanoparticle dimers}
\label{app:dimer}
Here we provide details of the calculation of the radiative frequency shifts in a 
homogeneous metallic nanoparticle dimer (see Fig.\ \ref{fig:sketch_dimer}), where the LSPs on each nanoparticle with (bare) resonance frequency $\omega_0$ 
are coupled through the near-field dipole-dipole interaction.

\subsection{Hamiltonian of the system}
The total Hamiltonian of the plasmonic dimer coupled to vacuum electromagnetic field modes reads as 
\begin{equation}
\label{eqS:H}
H=H_{\mathrm{pl}} + H_\mathrm{ph} + H_{\mathrm{pl}\textrm{-}\mathrm{ph}}.
\end{equation}
The plasmonic Hamiltonian describing near-field coupled LSPs is defined in the Coulomb gauge \cite{cohen, craig}
as (for details, see Refs.\ \cite{Brandstetter2015, Brandstetter2016})
\begin{align}
\label{eq:Hamdimer}
 H_{\mathrm{pl}} =&\; \hbar \omega_0 \sum_{n=1}^{2} \sum_{\sigma = x, y, z} {b_{n}^{\sigma}}^{\dagger} b_{n}^{\sigma} 
 \nonumber\\
&+ \hbar \Omega \sum_{\sigma = x, y, z} \eta_{\sigma} \left( b_{1}^{\sigma} + {b_{1}^{\sigma}}^{\dagger} \right) \left( b_{2}^{\sigma} + {b_{2}^{\sigma}}^{\dagger} \right),
\end{align}
where $\Omega = (\omega_0/2)({a}/{d})^3$ ($a$ is the nanoparticle radius and $d$ is the center-to-center interparticle distance; see Fig.\ \ref{fig:sketch_dimer}) and $\eta_{x,y} = 1$ ($\eta_{z} = -2$) for the transverse (longitudinal) modes. The bosonic operator $b_{n}^{\sigma}$ 
(${b_{n}^{\sigma}}^{\dagger}$) annihilates (creates) an LSP with polarization $\sigma$ on nanoparticle $n$. The quadratic Hamiltonian \eqref{eq:Hamdimer} is diagonalized by a Bogoliubov transformation, yielding Eq.\ \eqref{eq:pldimer}.
The bosonic operators in the latter equation are defined by
\begin{equation}
\label{eqS:Bog}
 B_{\tau}^{\sigma} = \sum_{n=1}^2 \left(u_{n \tau}^{\sigma} b_{n}^{\sigma} + \bar{u}_{n \tau}^{\sigma} {b_{n}^{\sigma}}^{\dagger}\right),
 \end{equation}
 where
 \begin{subequations}
 \label{eq:Bogcoeff}
\begin{align}
 u_{1 \tau}^{\sigma} &= \frac{\omega_{\tau}^{\sigma} + \omega_0}{2\sqrt{2 \omega_0 \omega_{\tau}^{\sigma}}},
 \\
 \bar{u}_{1 \tau}^{\sigma} &= \frac{\omega_{\tau}^{\sigma} - \omega_0}{2\sqrt{2 \omega_0 \omega_{\tau}^{\sigma}}},
\\
u_{2 \tau}^{\sigma} &= \tau\, \mathrm{sgn} \{ \eta_{\sigma} \} \frac{\omega_{\tau}^{\sigma} + \omega_0}{2\sqrt{2 \omega_0 \omega_{\tau}^{\sigma}}},
\\
\bar{u}_{2 \tau}^{\sigma} &= \tau\, \mathrm{sgn} \{ \eta_{\sigma} \}\frac{\omega_{\tau}^{\sigma} - \omega_0}{2\sqrt{2 \omega_0 \omega_{\tau}^{\sigma}}},
\end{align}
\end{subequations}
 while the inverse transformation reads as 
 $b_{n}^{\sigma} = \sum_{\tau=\pm}( u_{n \tau}^{\sigma} B_{\tau}^{\sigma} - \bar{u}_{n \tau}^{\sigma}  {B_{\tau}^{\sigma}}^{\dagger})$.
The operator $B_{\tau}^{\sigma}$ (${B_{\tau}^{\sigma}}^\dagger$) acts on an eigenstate $|n_\tau^\sigma\rangle$ of the Hamiltonian \eqref{eq:pldimer} representing a hybridized plasmon with polarization $\sigma$ and eigenenergy $\hbar\omega_\tau^\sigma$ as $B_{\tau}^{\sigma}|n_\tau^\sigma\rangle = \sqrt{n_\tau^\sigma}|n_\tau^\sigma-1\rangle$ 
(${B_{\tau}^{\sigma}}^\dagger|n_\tau^\sigma\rangle = \sqrt{n_\tau^\sigma+1}|n_\tau^\sigma+1\rangle$), 
with $n_\tau^\sigma$ a non-negative integer.

The photonic environment in Eq.\ \eqref{eqS:H} is described by the Hamiltonian \eqref{eq:H_ph}, while
the plasmon-photon coupling Hamiltonian is given in the long-wavelength limit ($k_0a\ll1$) by
\begin{equation}
\label{eqS:HamCoupling}
 H_{\mathrm{pl}\textrm{-}\mathrm{ph}} = \frac{e}{m_{\mathrm{e}}} \sum_{n=1}^{2} \mathbf{\Pi}_n \cdot \mathbf{A} (\mathbf{R}_n) + \frac{N_{\mathrm{e}} e^2}{2 m_{\mathrm{e}}} \sum_{n=1}^{2} \mathbf{A}^2 (\mathbf{R}_n), 
\end{equation}
where $\mathbf{R}_n$ corresponds to the location of the center of nanoparticle $n$, with $\mathbf{R}_2-\mathbf{R}_1=d\,\hat z$ (see Fig.\ \ref{fig:sketch_dimer}). 
In Eq.~\eqref{eqS:HamCoupling}, 
\begin{equation}
\mathbf{\Pi}_n=\mathrm{i}\sqrt{\frac{N_\mathrm{e}m_\mathrm{e}\hbar\omega_0}{2}}\sum_{\sigma=x,y,z} ({b_n^\sigma}^\dagger-b_n^\sigma)\hat\sigma
\end{equation}
corresponds to the momentum associated with the LSPs in nanoparticle $n$, while 
the vector potential $\mathbf{A}$ is given by Eq.~\eqref{eq:A}.
In terms of the Bogoliubov operators \eqref{eqS:Bog}, the plasmon-photon coupling \eqref{eqS:HamCoupling} hence reads 
\begin{align}
\label{eqS:H_c}
H_{\mathrm{pl}\textrm{-}\mathrm{ph}} =&\;
\mathrm{i}\hbar\sum_{n=1}^2\sum_{\sigma, \tau}
\sum_{\mathbf{k}, \hat\lambda_{\mathbf{k}}}
\sqrt{\frac{\pi\omega_0^3 a^3}{\mathcal{V}\omega_\mathbf{k}}}
\hat\sigma\cdot\lambda_{\mathbf{k}}
\left(u_{n\tau}^\sigma+\bar u_{n\tau}^\sigma\right)
\nonumber\\
&\times\left({B_\tau^\sigma}^\dagger-B_\tau^\sigma\right)
\left(
a_\mathbf{k}^{\hat\lambda_{\mathbf{k}}}\mathrm{e}^{\mathrm{i}\mathbf{k}\cdot\mathbf{R}_n}
+{a_\mathbf{k}^{\hat\lambda_{\mathbf{k}}}}^\dagger\mathrm{e}^{-\mathrm{i}\mathbf{k}\cdot\mathbf{R}_n}
\right)
\nonumber\\
&+\pi\hbar\omega_0^2\frac{a^3}{\mathcal{V}}
\sum_{n=1}^2
\sum_{\substack{\mathbf{k}, \hat\lambda_{\mathbf{k}}\\\mathbf{k}', \hat\lambda'_{\mathbf{k}'}}}
\frac{\lambda_{\mathbf{k}}\cdot\lambda_{\mathbf{k}'}}{\sqrt{\omega_\mathbf{k}\omega_{\mathbf{k}'}}}
\nonumber\\
&\times
\left(
a_\mathbf{k}^{\hat\lambda_{\mathbf{k}}}\mathrm{e}^{\mathrm{i}\mathbf{k}\cdot\mathbf{R}_n}
+{a_\mathbf{k}^{\hat\lambda_{\mathbf{k}}}}^\dagger\mathrm{e}^{-\mathrm{i}\mathbf{k}\cdot\mathbf{R}_n}
\right)
\nonumber\\
&\times
\left(
a_{\mathbf{k}'}^{\hat\lambda_{\mathbf{k}'}}\mathrm{e}^{\mathrm{i}\mathbf{k}'\cdot\mathbf{R}_n}
+{a_{\mathbf{k}'}^{\hat\lambda_{\mathbf{k}'}}}^\dagger\mathrm{e}^{-\mathrm{i}\mathbf{k}'\cdot\mathbf{R}_n}
\right).
\end{align}

\subsection{Nondegenerate perturbation theory}

Carrying out a perturbative calculation up to second order in the coupling Hamiltonian \eqref{eqS:H_c}, the plasmonic energy levels read
$E_{n_{\tau}^\sigma} =E_{n_{\tau}^\sigma}^{(0)} + E_{n_{\tau}^\sigma}^{(1)} + E_{n_{\tau}^\sigma}^{(2)}$, where
the zeroth-order result is 
$E_{n_{\tau}^\sigma}^{(0)} = n_{\tau}^{\sigma} \hbar \omega_{\tau}^{\sigma}$.
The first-order correction
\begin{equation}
E_{n_\tau^\sigma}^{(1)} =4\pi\hbar\omega_0^2\frac{a^3}{\mathcal{V}}\sum_\mathbf{k}\frac{1}{\omega_\mathbf{k}}
\end{equation}
 is twice as much as the single-nanoparticle result [see Eq.\ \eqref{eq:E1}]. It also represents a global 
 energy shift experienced by all the hybridized-mode levels and hence does not contribute to a renormalization of the resonance 
 frequencies. 
The second-order correction is 
\begin{align}
\label{eq:Dimersecondorder}
E_{n_{\tau}^\sigma}^{(2)} =&\; \pi \hbar \omega_0^2 \omega_{\tau}^{\sigma} \frac{a^3}{\mathcal{V}} \sum_{\mathbf{k}, \hat{\lambda}_{\mathbf{k}}} \frac{|\hat{\sigma} \cdot \hat{\lambda}_{\mathbf{k}} |^2}{\omega_{\mathbf{k}}} \frac{ (2 n_{\tau}^{\sigma} + 1) \omega_{\mathbf{k}}-\omega_{\tau}^{\sigma}}{{\omega_{\tau}^{\sigma}}^2 - \omega_{\mathbf{k}}^2} 
\nonumber\\
&\times
\left[ 1 + \tau \mathrm{sgn} \{ \eta_{\sigma} \} \cos (k_z d)\right],
\end{align}
where $k_z=\mathbf{k}\cdot\hat{z}$, and where the summation excludes the term for which $\omega_\mathbf{k}=\omega_{\tau}^{\sigma}$. The same renormalization procedure \cite{Bethe1947, Milonni1994} as in the case of a single nanoparticle (see Appendix \ref{app:1NP}) yields
\begin{align}
\label{eq:DimerRenorm}
 \bar{E}_{n_{\tau}^\sigma}^{(2)} =&\; \pi\hbar \omega_0^2 {\omega_{\tau}^{\sigma}}^2 \frac{a^3}{\mathcal{V}} \sum_{\mathbf{k}, \hat{\lambda}_{\mathbf{k}}} \frac{|\hat{\sigma} \cdot \hat{\lambda}_{\mathbf{k}} |^2}{\omega_{\mathbf{k}}^2} \frac{ (2 n_{\tau}^{\sigma} + 1) \omega_{\tau}^{\sigma}- \omega_{\mathbf{k}}}{{\omega_{\tau}^{\sigma}}^2 - \omega_{\mathbf{k}}^2} 
 \nonumber\\
 &\times
 \left[ 1 + \tau \mathrm{sgn} \{ \eta_{\sigma} \} \cos (k_z d)\right].
\end{align}
The renormalized frequency difference between successive hybridized plasmonic levels is then $ \tilde\omega_{\tau}^{\sigma} = \omega_{\tau}^{\sigma} + \delta_{\tau}^{\sigma}$, 
where the radiative shift reads
\begin{equation}
\label{eqS:delta}
\delta_{\tau}^{\sigma}=
2\pi  \omega_0^2 {\omega_{\tau}^{\sigma}}^3 \frac{a^3}{\mathcal{V}} \sum_{\mathbf{k}, \hat{\lambda}_{\mathbf{k}}} \frac{|\hat{\sigma} \cdot \hat{\lambda}_{\mathbf{k}} |^2}{\omega_{\mathbf{k}}^2} \frac{1 + \tau \mathrm{sgn} \{ \eta_{\sigma} \} \cos (k_z d)}{{\omega_{\tau}^{\sigma}}^2 - \omega_{\mathbf{k}}^2}. 
\end{equation}

\subsection{Explicit evaluation of $\delta_{\tau}^{\sigma}$}
Replacing the summation over photon momenta in Eq.~\eqref{eqS:delta} by a three-dimensional principal-value integral (up to the ultraviolet cutoff $k_\mathrm{c}$ which is of order $1/a$) and using Eq.\ \eqref{eqS:sum_pol} yields
\begin{align}
\label{eqS:theta}
\delta_\tau^\sigma=&\;\frac{\omega_0^2{\omega_\tau^\sigma}^2a^3}{(2\pi c)^2}
\mathcal{P}\int_0^{k_\mathrm{c}}\frac{\mathrm{d}k}{{\omega_\tau^\sigma}^2-(ck)^2}
\nonumber\\
&\times
\int_0^\pi\mathrm{d}\theta\sin{\theta}\left[1+\tau\, \mathrm{sgn} \{\eta_{\sigma}\} \cos{(kd\cos{\theta})}\right]
\nonumber\\
&\times
\int_0^{2\pi}\mathrm{d}\varphi[1-(\hat k\cdot\hat\sigma)^2].
\end{align}
The integral over $\varphi$ is given in Eq.\ \eqref{eqS:int_phi} so that the integral over $\theta$ in Eq.\ \eqref{eqS:theta} yields
\begin{align}
\delta_\tau^\sigma=&\;\frac{2\omega_0^2{\omega_\tau^\sigma}^3a^3}{3\pi c^3}
\mathcal{P}\int_0^{\omega_\mathrm{c}}\frac{\mathrm{d}\omega}{{\omega_\tau^\sigma}^2-\omega^2}
\nonumber\\
&+\tau\frac{|\eta_\sigma|\omega_0^2{\omega_\tau^\sigma}^3a^3}{2\pi c^2 d}
\mathcal{P}\int_0^{\omega_\mathrm{c}}
\frac{\mathrm{d}\omega}{\omega\left({\omega_\tau^\sigma}^2-\omega^2\right)}
\nonumber\\
&\times
\left\{
\left[
1+\mathrm{sgn}\{\eta_\sigma\}-2\left(\frac{c}{\omega d}\right)^2
\right]\sin{\left(\frac{\omega d}{c}\right)}\right.
\nonumber\\
&+\left.\frac{2c}{\omega d}\cos{\left(\frac{\omega d}{c}\right)}
\right\}.
\end{align}
A lengthy, but straightforward calculation then gives the result of Eq.\ \eqref{eq:shiftdimer}.

\section{Derivation of Eq.\ \eqref{eq:beauty}}
\label{app:beauty}
In this Appendix, we provide details of the calculation for the ratio 
\begin{equation}
\label{eq:ratio_app}
\frac{\Delta\tilde\omega^z}{\Delta\tilde\omega^{x,y}}=\frac{\omega_+^z-\omega_-^z+\delta_+^z-\delta_-^z}{\omega_+^{x,y}-\omega_-^{x,y}+\delta_+^{x,y}-\delta_-^{x,y}},
\end{equation}
leading to Eq.\ \eqref{eq:beauty}. Here, the bare resonance frequencies of the hybridized plasmonic modes $\omega_\tau^\sigma$ and their associated radiative frequency shifts $\delta_\tau^\sigma$ are given by Eqs.\ \eqref{eq:omega_bare} and \eqref{eq:shiftdimer}, respectively.

With the cutoff $\omega_\mathrm{c}=c/a$, in the limit where $\Omega/\omega_0\ll 1$, $k_0a\ll1$, and $k_0d\ll1$, Eq.\ \eqref{eq:shiftdimer} reduces to 
\begin{align}
\label{eqS:approx_shift}
\delta_\tau^\sigma\simeq&\;
\delta_0+\tau\frac{3|\eta_\sigma|}{8}\delta_0
\bigg\{
4\left(\frac{a}{d}\right)^3
\nonumber\\
&+\left[\left(\frac{a}{d}\right)^2+\frac 12+\mathrm{sgn}\{\eta_\sigma\}\right]\cos{\left(\frac da\right)}
\nonumber\\
&+\frac ad \left[-\left(\frac{a}{d}\right)^2+\frac 12+\mathrm{sgn}\{\eta_\sigma\}\right]\sin{\left(\frac da\right)}
\nonumber\\
&+\frac da\left(\frac 12+\mathrm{sgn}\{\eta_\sigma\}\right)\mathrm{Si}\left(\frac da\right)
\bigg\},
\end{align}
where $\delta_0\simeq2\omega_0(k_0a)^4/3\pi$ corresponds to the single-particle radiative shift derived in Sec.\ \ref{sec:1NP}.
Using the expression \eqref{eqS:approx_shift}, we then obtain to leading order in $a/d$
\begin{subequations}
\label{eq:deltadelta}
\begin{align}
\delta_+^z-\delta_-^z&\simeq-\frac{\omega_0}{4}(k_0a)^4\frac da,\\
\delta_+^{x,y}-\delta_-^{x,y}&\simeq\frac{3\omega_0}{8}(k_0a)^4\frac da, 
\end{align}
\end{subequations}
so that the ratio
$\Delta\tilde\omega^z/\Delta\tilde\omega^{x,y}$ reduces to Eq.\ \eqref{eq:beauty}.

As discussed in the main text, it is important to check the robustness of the result \eqref{eq:beauty} against multipolar interactions of the LSPs with 
the photonic environment beyond the dipolar approximation in Eq.\ \eqref{eq:plphsingle}. 
To leading order in $k_0a\ll1$, the radiative shifts in Eqs.\ 
\eqref{eq:shiftsingle} and \eqref{eq:shiftdimer} scale as $(k_0a)^4$. One can then expect that the next leading-order correction to these results goes, at least, as $(k_0a)^5$. Taking into account such corrections would lead to a modification of Eq.\ \eqref{eq:deltadelta} that reads
\begin{subequations}
\label{eq:deltadelta_2}
\begin{align}
\delta_+^z-\delta_-^z&\simeq-\frac{\omega_0}{4}(k_0a)^4\frac da+\alpha^z \omega_0 (k_0a)^5,\\
\delta_+^{x,y}-\delta_-^{x,y}&\simeq\frac{3\omega_0}{8}(k_0a)^4\frac da+\alpha^{x,y} \omega_0 (k_0a)^5, 
\end{align}
\end{subequations}
where $\alpha^z$ and $\alpha^{x,y}$ are some constants of the order of unity. Incorporating the above expressions in Eq.\ \eqref{eq:ratio_app} yields
\begin{equation}
\label{eq:ratio_nonu}
\frac{\Delta\tilde\omega^z}{\Delta\tilde\omega^{x,y}}\simeq 2-(k_0d)^4+\left(\alpha^z-2\alpha^{x,y}\right)(k_0a)^2 (k_0d)^3. 
\end{equation}
The last term of Eq.\ \eqref{eq:ratio_nonu} represents an $a$-dependent, nonuniversal correction to the result \eqref{eq:beauty}. However, such a term is negligible for the parameter regime which we consider in the present work, as $(k_0a)^2\ll k_0d$.

\section{Radiative frequency shifts in heterogeneous nanoparticle dimers}
\label{app:hetero}
In this Appendix, we consider a heterogeneous dimer where the two nanoparticles have different resonance frequencies $\omega_1$ and $\omega_2$, which may be due to small variations in their size and/or shape. We present the effective model describing such a situation and the explicit form of the frequency shifts $\delta_{\tau}^{\sigma}$. 
We also provide approximate expressions for the crucial dimensionless ratios $\Delta \omega^z / \Delta \omega^{x, y}$ (without the photonic environment) and $\Delta \tilde{\omega}^z / \Delta \tilde{\omega}^{x, y}$ (with the photonic environment) for this heterogeneous case.

The effective plasmonic Hamiltonian for a heterogeneous nanoparticle dimer reads as
\begin{align}
\label{eq:Ham_dimer_hetero}
 H_{\mathrm{pl}} =&\;  \sum_{n=1}^{2} \sum_{\sigma = x, y, z} \hbar \omega_n {b_{n}^{\sigma}}^{\dagger} b_{n}^{\sigma} 
 \nonumber\\
&+ \hbar \Omega \sum_{\sigma = x, y, z} \eta_{\sigma} \left( b_{1}^{\sigma} 
+ {b_{1}^{\sigma}}^{\dagger} \right) \left( b_{2}^{\sigma} + {b_{2}^{\sigma}}^{\dagger} \right),
\end{align}
where the coupling constant is $\Omega = (\sqrt{\omega_1 \omega_2}/2)({\bar{a}}/{d})^3$, the average nanoparticle radius is $\bar{a}$ and the center-to-center interparticle distance is $d$. The Hamiltonian \eqref{eq:Ham_dimer_hetero} can be diagonalized exactly, yielding the coupled mode eigenfrequencies~\cite{Brandstetter2015, Brandstetter2016}
\begin{equation}
\label{eq:Eigen_dimer_hetero}
\omega_{\tau}^{\sigma} =  \sqrt{ \frac{\omega_1^2 + \omega_2^2}{2} + \tau \sqrt{ 4 \eta_{\sigma}^2 \Omega^2 \omega_1 \omega_2 + \left( \frac{\omega_1^2-\omega_2^2}{2} \right)^2 }}.
\end{equation}
Introducing the frequency difference $\delta \omega = \omega_1 - \omega_2$ and the average resonance frequency $\bar{\omega} = \left(\omega_1 + \omega_2\right)/2$, 
the ratio $\Delta \omega^z / \Delta \omega^{x, y}$ (neglecting coupling to the photonic modes) reads
\begin{equation}
\label{eq:Ratio_hetero} 
 \frac{\Delta \omega^z}{\Delta \omega^{x, y}} \simeq 2 \sqrt{\frac{ \Omega^2 +  {\delta \omega^2}/{16} }{ \Omega^2 + \delta \omega^2/4 }}
\end{equation}
in the limit $\delta \omega / \bar{\omega} \ll 1$ and $\Omega / \bar{\omega} \ll 1$, 
with $\Omega \simeq \left( \bar{\omega} /2 \right) \left( \bar{a} / d \right)^3$. 
In the small detuning regime where $\delta\omega\ll\Omega$, the expression above reduces to
\begin{equation}
\label{eq:Ratio_approx_hetero} 
 \frac{\Delta \omega^z}{\Delta \omega^{x, y}} \simeq 2 - \frac{3}{16} \left(\frac{\delta \omega}{\Omega}\right)^2,
\end{equation}
which recovers the result given in Sec.\ \ref{sec:dimer} for a homogeneous dimer up to the quadratic correction in $\delta \omega / \Omega$.

The expression for the radiative frequency shifts for heterogeneous dimers, which is obtained analogous to the case of homogeneous dimers (see Appendix \ref{app:dimer}), reads
\begin{align}
\label{eq:Shift_hetero}
\delta_{\tau}^{\sigma} =&\;
 \frac{\bar{a}^3}{3 \pi c^3} \frac{{\omega_{\tau}^{\sigma}}^2}{2{\omega_{\tau}^{\sigma}}^2 - \omega_1^2 - \omega_2^2} 
 \nonumber\\
&\times \left[ \omega_1^2 \left( {\omega_{\tau}^{\sigma}}^2 - \omega_2^2 \right) + \omega_2^2 \left( {\omega_{\tau}^{\sigma}}^2 - \omega_1^2 \right) \right]
\nonumber\\
&\times \ln{ \left( \frac{ \omega_\mathrm{c} / \omega_{\tau}^{\sigma} + 1 }{\omega_\mathrm{c} / \omega_{\tau}^{\sigma} - 1} \right)}
\nonumber\\
 &
 + \tau
 \frac{ |\eta_{\sigma}|\omega_1\omega_2}{\pi\omega_{\tau}^{\sigma}} \left( \frac{\bar{a}}{d}\right)^3 \frac{\sqrt{{\omega_{\tau}^{\sigma}}^2 - \omega_1^2} \sqrt{{\omega_{\tau}^{\sigma}}^2 - \omega_2^2}}{2{\omega_{\tau}^{\sigma}}^2 - \omega_1^2 - \omega_2^2}  g_{\tau}^{\sigma},
\end{align}
where $g_{\tau}^{\sigma}$ is defined in Eq.\ \eqref{eq:superlong}. Expanding the above exact result in the limit $\delta \omega / \bar{\omega} \ll 1$, $\Omega / \bar{\omega} \ll 1$, $\bar{k} \bar{a} \ll 1$ and $\bar{k} d \ll 1$ (with $\bar{k} = \bar{\omega}/c$), leads to the following compact, approximate expression for the ratio $\Delta \tilde{\omega}^z / \Delta \tilde{\omega}^{x, y}$ (including coupling to the photonic modes) in the regime $\delta \omega / \Omega \ll 1$,
\begin{equation}
\label{eq:Approx_hetero}
 \frac{\Delta \tilde{\omega}^z}{\Delta \tilde{\omega}^{x, y}} \simeq 2 - \left( \bar{k} d\right)^4 - \frac{3}{16}\left( \frac{\delta \omega}{\Omega}\right)^2.
\end{equation}
This expression recovers the corresponding result \eqref{eq:beauty} for a homogeneous dimer, up to the quadratic correction in $\delta \omega / \Omega$. Thus small differences in the LSP resonance frequencies of the nanoparticles comprising the dimer do not rule out our proposed experiment to unambiguously detect radiative frequency shifts.


\end{document}